\documentclass[12pt]{article}
\makeatletter
\def\ps@pprintTitle{%
	\let\@oddhead\@empty
	\let\@evenhead\@empty
	\def\@oddfoot{}%
	\let\@evenfoot\@oddfoot}
\makeatother

\usepackage[utf8]{inputenc}
\usepackage{fontenc}[t1]

\usepackage{comment}   
\usepackage{geometry}                
\usepackage{amssymb}
\usepackage{amsmath}
\usepackage{amsthm}
\usepackage{bigints}
\usepackage{adjustbox}

\usepackage{lineno}

\usepackage{graphicx}
\usepackage{epstopdf}
\newcommand{\field}[1]{\mathbb{#1}}
\newcommand {\R}        {\field{R} }

\newcommand {\eps}      {\varepsilon}

\newcommand {\xt}{\tilde x}
\newcommand {\yt}{\tilde y}
\newcommand {\zt}{\tilde z}
\newcommand {\wt}{\tilde w}

\usepackage{color}
\usepackage{colordvi}
\definecolor{magenta}{rgb}{.5,0,.5}
\definecolor{black}{rgb}{1.0,1.0,1.0}
\definecolor{magenta}{rgb}{.1,0,.3}
\definecolor{gruen}{rgb}{0.2,0.5,.5}
\definecolor{light}{rgb}{ 0.992, 0.961,  0.902}
\definecolor{Tan}{rgb}{ 0.992, 0.9,  0.902}

\theoremstyle{plain} 

\newtheorem{prop}{Proposition}[section]

\newtheorem{theorem}[prop]{Theorem}

\newtheorem{cor}[prop]{Corollary}

\newcommand{\komment}[1]{}

\title{Life-History traits and the replicator equation}

\author{Johannes M\"uller\footnote{Center for Mathematics, Technische Universit\"at M\"unchen, 85748 Garching, Germany} 
	\footnote{Institute for Computational Biology, Helmholtz Center Munich, 85764 Neuherberg, Germany}, \ 
	Aur\'elien Tellier\footnote{Professorship for Population Genetics, Department of Life Science Systems, School of Life Sciences, Technische Universit\"at M\"unchen, 85354 Freising, Germany}} 

\begin{document}

\maketitle

\begin{abstract}
Due to the relevance for conservation biology, there is an increasing interest to extend evolutionary genomics models to plant, animal or microbial species. However, this requires to understand the effect of life-history traits absent in humans on genomic evolution. In this context, it is fundamentally of interest to generalize the replicator equation, which is at the heart of most population genomics models. However, as the inclusion of life-history traits generates models with a large state space, the analysis becomes involving. We focus, here, on quiescence and seed banks, two features common to many plant, invertebrate and microbial species. We develop a method to obtain a low-dimensional replicator equation in the context of evolutionary game theory, based on two assumptions: (1) the life-history traits are {\it per se} neutral, and (2) frequency-dependent selection is weak. We use the results to investigate the evolution and maintenance of cooperation based on the Prisoner's dilemma. We first consider the generalized replicator equation, and then refine the investigation using adaptive dynamics. It turns out that, depending on the structure and timing of the quiescence/dormancy life-history trait, cooperation in a homogeneous population can be stabilized. We finally discuss and highlight the relevance of these results for plant, invertebrate and microbial communities.  
\end{abstract}

Keywords: Replicator equation, life-history traits, seed bank, quiescence, persister, Prisoner's dilemma, adaptive dynamics.\par\medskip

\section{Introduction}

One of the main goals of population genetics/genomics is to understand the evolutionary dynamics of populations/species under the influence of several neutral and selective evolutionary forces. The methods range from rather abstract approaches such as evolutionary game theory~\cite{Hofbauer1998} to very specific and realistic models, \textit{e.g.} investigating the cooperative traits of bacterial communities  \cite{West2007a,Hoesel2019} to only name one example. However, most investigations focus on one single trait, as in the core theory of adaptive dynamics~\cite{Metz1995,DiekmannIntro}. In recent time it becomes more and more clear that particular life-history traits can be game-changer~\cite{McNamara2013}. For example, cannibalism can only be understood as a favourable strategy if age structure is taken into account~\cite{Getto2005,Garay2016}, or some life-history traits might shift evolutionary stable strategies under frequency-dependent selection away from the point of maximum mean fitness~\cite{Day1996}. There are several recent attempts to reformulate the replicator equation, which is a most powerful tool in the context of evolutionary game theory, in an age-structured setting~\cite{Lessard2017,Argasinski2021,Li2015}. All these approaches allow the life-history traits and the frequency-dependent selection to run on similar time scales, and to have similar strength. Accordingly, the state space of the replicator-equation becomes large, so that starting with a model based on continuous age structure a coupled PDE-ODE system is obtained, otherwise if the age structure is formulated in a discrete way, a high-dimensional ODE is obtained. We emphasize that the advantage of this approach is its generality, as no further assumptions are necessary. A disadvantage lies nevertheless in the high dimension of the resulting replicator-equation, and the high technical effort necessary for a rigorous analysis. However, if we are to investigate the evolution of genomes or cooperation in plant, invertebrate or microbial communities, it is necessary to take into account common life-history traits, such as age-structure, quiescence or dormancy.   \par\medskip 

In the present paper, we use a more restricted setting, that allows to simplify the results. In the case of two competing populations, we obtain a one-dimensional ODE. We assume here that generation overlap or age structure is due to quiescence or seed dormancy. Quiescence and dormancy appear in different contexts. One prototypical biological system with quiescence are bacteria that exhibit a persister state~\cite{Balaban2004}. Persister cells minimize their metabolism, such that \textit{e.g.}\ antibiotics cannot harm them, which is obviously a major issue in medicine. Plants represent another important biological system with a dormant state, as many flowering plants develop so-called seed banks. Indeed, seeds have the ability to remain dormant for a long time in the soil. We note that the two cases are slightly different, though both assume that the persister/quiescent/dormant state does not divide. On the one hand a given bacterial individual can switch between quiescence and active state and is active right after cell division. On the other hand, plants always reproduce via the dormant state. Recently, it has been shown that quiescence/dormant has manifold effects on genetic heterogeneity within populations~\cite{kaj2001,Blath2013,tellier2009}, on the speed and strength of selection~\cite{Koopmann2017,Heinrich2017}, on the interplay with variable environment~\cite{Blath2020,Blath2021,Kussell2005,Mueller2013}, and on co-evolution between hosts and parasites~\cite{verin2018host}. We redirect the interested reader to two recent review articles~\cite{Lennon2021, tellier2019}.\par\medskip 

In the current study, we focus on the effect of different timescales in the quiescent/dormant phase, and the co-evolution of these time scales with a second trait that superimposes frequency-dependent selection (Section~2). We formulate this second trait in the spirit of evolutionary game theory, extending our previous work~\cite{Sellinger2019}.  While in~\cite{Sellinger2019}, exponentially growing populations have been considered, in the present we focus on a population of constant size. Two main ingredients allow us to reduce the system describing the competition of two populations to a one-dimensional replicator equation. First, we assume that the life-history traits without frequency-dependent selection are neutral, and second, we assume that the frequency-dependent selection is weak. The combination of the two assumptions allows to use singular perturbation theory~\cite{OMalley2012} to obtain a simplifying approximation of the dynamics. Interestingly, we find that the shape of the resulting generalized replicator equation depends particularly on the model-assumptions for the life-history traits, namely a difference is observed between quiescence and seed bank.\par\medskip 

As an application, we consider in Section~3 the prisoner's dilemma to investigate the co-evolution of time scales and cooperation. Generally, we note that mechanisms that stabilize cooperation, that is its evolution and maintenance, are of special interest in evolutionary biology. In homogeneous populations, without additional structure, cooperation mostly gets lost under evolutionary pressure, a result which is often termed as the tragedy of the commons~\cite{West2007a, AsfahlSchuster2016}. If, by some mechanisms, cooperators are more likely to interact with other cooperators, then cooperation might persist. The most common idea why this assortment can take place is to assume the existence of spatial  structuring~\cite{price,okasha2009,AllenNowak2014}. Another idea is, for example, that cooperators have the ability to recognize other cooperators by so-called ``greenbeard''-genes~\cite{WestGardner2010}. This approach requires thus the co-evolution of cooperation and greenbeardness. In the same spirit, we consider here the co-evolution of cooperation and quiescence/dormancy. We show that this co-evolution between traits can lead under appropriate conditions to an evolutionary stable cooperation. Interestingly enough, when the population size is constant, cooperation can be stable only under seed banking, and not for bacterial population with persister/quiescent state. This finding contradicts our previous results for exponentially growing populations~\cite{Sellinger2019}, where persister cells could also stabilize cooperation. In the last section we thus discuss our results and this apparent paradox.

\section{Models and analysis}
\begin{figure}[h!]
	\begin{center}
		\includegraphics[width=12cm]{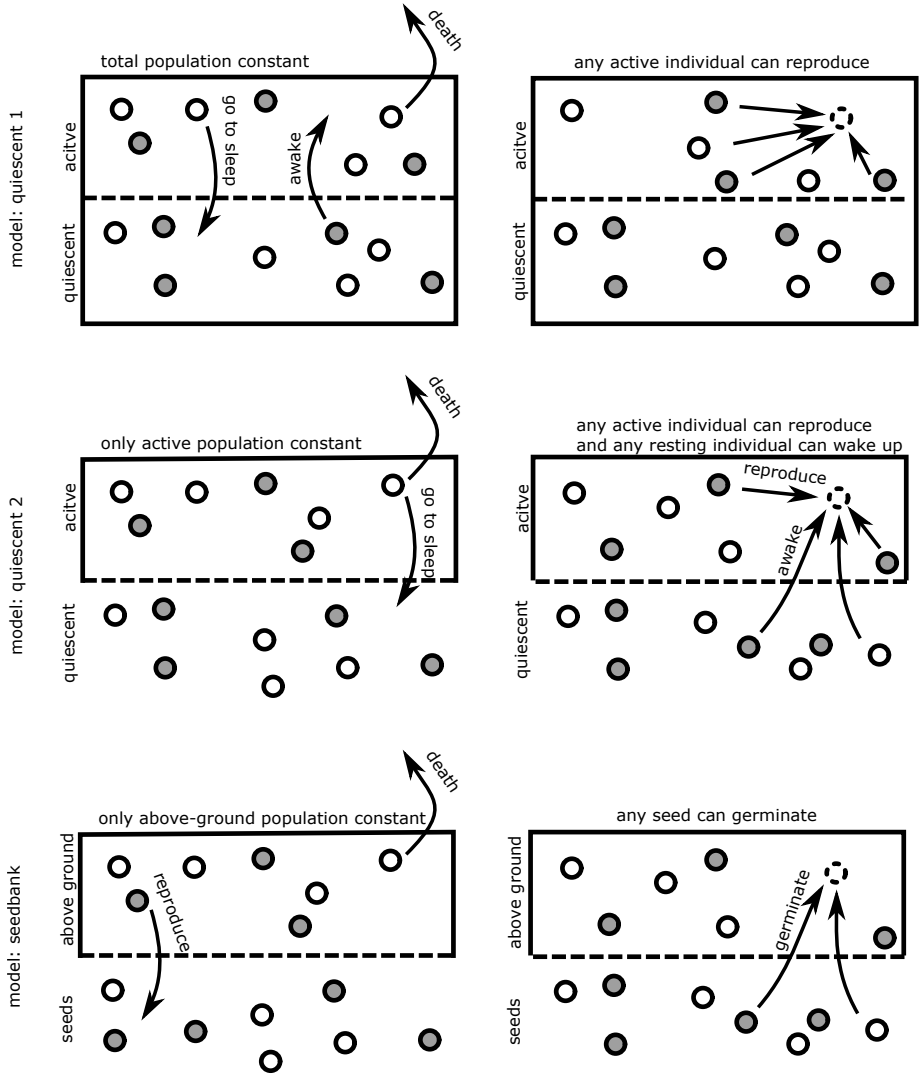}
	\end{center}
	\caption{Scheme of the three models: Quiescence I, quiescence II, and seed bank.}\label{modStuct}
\end{figure}

We assume a time scale separation between ecological and evolutionary processes, as it is the basis for adaptive dynamics~\cite{Metz1995,DiekmannIntro}. At the ecological level, we investigate the competition of two types of individuals, differing in the degree of cooperation as well as in the time scale for quiescence/dormancy. As it is rather usual in population genetics, we keep the total population size constant, addressing  an ecosystem at its carrying capacity. To investigate the effect of different quiescence/dormant mechanisms, we develop several variants of our model: persister cells in bacterial communities are different to seed banks in plant populations. The main difference is the reproduction -- while in bacteria the offspring is active, reproduction in plants yields quiescent/dormant individuals (seeds). For persister cells, the assumption of a constant population size can be interpreted in two ways: either the total (model Quiescence I) or only the active population (model Quiescence II) is constant (see Fig.~\ref{modStuct}). For seed banks, only the above-ground population is kept constant. In consequence, we build three ecological models: one seed bank model, and two persister models. Though we investigate deterministic (ODE) models, we start off with a stochastic model formulation, as we find it simpler to formulate models with a constant population size in an individual-based setting. Constructing the deterministic limit yields the appropriate ODE models.\\
We then augment the three models with weak frequency-dependent selection in the spirit of evolutionary game theory, and then use singular perturbation theory to work out the generalized replicator equation.

\subsection{Quiescence I -- total population is constant}
We consider a constant population of size $N$. Individuals are of type~A or type~B, and both types can fall in a quiescent state at rate $\eta_A$ ($\eta_B$), resp.\ reactivate themselves at rate $\zeta_A$ ($\zeta_B$). Only active individuals die (at rate $\mu$). If an individual dies, it is immediately replaced by a newborn individual to keep the population size constant. Each of the active individuals have the same probability to reproduce. Let $X_t$ ($Y_t$) denote the active (quiescent) population of type A, and  $Z_t$ ($W_t$) the corresponding sub-populations of type B.
\par\medskip 

\begin{table}
\begin{tabular}{lll}
	\hline
	Type & Transition & Rate\\
	\hline
	death B \& birth type A 
	& $(X_t,Y_t,Z_t, W_t)\rightarrow(X_t+1,Y_t,Z_t-1, W_t)$
	&$\mu Z_t\,\frac{X_t}{X_t+Z_t}$\\
	death A \& birth type B 
	& $(X_t,Y_t,Z_t, W_t)\rightarrow(X_t-1,Y_t,Z_t+1, W_t)$
	&$\mu X_t\,\frac{Z_t}{X_t+Z_t}$\\
	go quiescent type A 
	& $(X_t,Y_t,Z_t, W_t)\rightarrow(X_t-1,Y_t+1,Z_t, W_t)$
	&$\eta_A X_t$\\
	wake up type A 
	& $(X_t,Y_t,Z_t, W_t)\rightarrow(X_t+1,Y_t-1,Z_t, W_t)$
	&$\zeta_A Y_t$\\
	go quiescent type B 
	& $(X_t,Y_t,Z_t, W_t)\rightarrow(X_t,Y_t,Z_t-1, W_t+1)$
	&$\eta_B Z_t$\\
	wake up type B 
	& $(X_t,Y_t,Z_t, W_t)\rightarrow(X_t,Y_t,Z_t+1, W_t-1)$
	&$\zeta_B W_t$\\
	\hline 
\end{tabular}
\caption{Transitions for the quiescence model I}\label{qmodItransitions}
\end{table}

We obtain the system of ODE's, which is the deterministic counterpart of the stochastic model described in table~\ref{qmodItransitions}.
\begin{eqnarray}
	\frac d {dt} x &=& 
	\mu z\frac{x}{x+z}-\mu x\frac{z}{x+z}-\eta_Ax+\zeta_A y,\\
	\frac d {dt} y &=& \eta_Ax-\zeta_A y,\\
	\frac d {dt} z &=& 
	-\mu z\frac{x}{x+z}+\mu x\frac{z}{x+z}
	- \eta_Bz+\zeta_B w,\\
	\frac d {dt} w &=& 
	\eta_Bz-\zeta_B w.
\end{eqnarray}
where $x+y+z+w=1$. 
Obviously, the non-linear terms cancel, and we are left with a linear model which possesses a line of stationary points,

$$
\left(\begin{array}{c}
	x\\y\\z\\w
\end{array}\right)
= 
\left(\begin{array}{c}
	u\frac{\zeta_A}{\zeta_A+\eta_A}\\
	u\frac{\eta_A}{\zeta_A+\eta_A}\\
	(1-u)\frac{\zeta_B}{\zeta_B+\eta_B}\\
	(1-u)\frac{\eta_B}{\zeta_B+\eta_B}
\end{array}\right),\qquad 
u\in[0,1].
$$
That is, the fitness of type-A and type-B individuals are exactly the same. We therefore call the model as being ``neutral'' with respect to the type, even though the parameters of type A are in general not identical with that of type B.

\subsubsection{Weak selection}
In order to incorporate weak frequency-dependent selection, we introduce pay-off-functions $g_A(x,z)$ and $g_B(x,z)$ for type A and type B, and modify the probabilities of the birth events. The augmented model becomes:
\begin{eqnarray}
	\frac d {dt} x &=& 
	\frac{\mu z(x+\eps g_A(x,z)\, x)}{x+z+\eps(g_A(x,z) x+g_B(x,z) z)}-\frac{\mu x(z+\eps g_B(x,z) z)}{x+z+\eps(g_A(x,z) x+g_B(x,z) z)}\\
	&&-\eta_Ax+\zeta_A y\nonumber, \\
	\frac d {dt} y &=& \eta_Ax-\zeta_A y,\\
	\frac d {dt} z &=& 
	\frac{-\,\mu z(x+\eps g_A(x,z) x)}{x+z+\eps(g_A(x,z) x+g_B(x,z) z)}+\frac{\mu x(z+\eps g_B(x,z) z)}{x+z+\eps(g_A(x,z) x+g_B(x,z) z)}\,\,\,\\
	&&- \eta_Bz+\zeta_B w\nonumber, \\
	\frac d {dt} w &=& 
	\eta_Bz-\zeta_B w.
\end{eqnarray}
Note that the total population size is still constant as
$ \frac d{dt} (x+y+z+w)=0$. 
It is sufficient to consider $x$, $y$, $w$. In the next equation, we use $z$ in the understanding that $z=1-x-y-w$, 
\begin{eqnarray*}
	\frac d {dt} x &=& 
	\frac{\eps \,\mu \,x\,z\,\,(\,g_A(x,z) - g_B(x,z)\,)}{x+z+\eps(g_A(x,z) x+g_B(x,z) (1-x-y-w) )},\\
	\frac d {dt} y &=& \eta_Ax-\zeta_A y,\\
	\frac d {dt} w &=& 
	\eta_B(1-x-y-w)-\zeta_B w.
\end{eqnarray*}
We furthermore introduce the fraction of type-A individuals $\tilde x=x+y$, and replace $x$ by 
$\tilde x-y$, and $z$ by $1-x-y-w=1-\tilde x-w$. In what follows, $x$ and $z$ are not independent variables as before, but only abbreviations for the indicated expression. Therewith, 
\begin{eqnarray}
	\frac d {dt} {\tilde x}
	&=& 
	\mu \eps \frac{(\,\,g_A(x,z)-g_B(x,z)\,\,)\,(\tilde x-y) \, (1-\tilde x-w)}
	{1-y-w+\eps(g_A(x,z) x+g_B(x,z) z)},\\
	\frac d {dt} y &=& \eta_A(\tilde x-y)-\zeta_A y,\\
	\frac d {dt} w &=& 
	\eta_B(1-\tilde x-w)-\zeta_B w.
\end{eqnarray}

The analysis based on singular perturbation theory is rather straightforward. If we take $\eps$ to zero, we find that the slow variable $\tilde x$ does not change, and the fast variables $y$ and $w$ satisfy a linear system, 
\begin{eqnarray*}
	\frac d {dt} \tilde x 
	&=&  0,\\
	\frac d {dt} y &=& \eta_A(\tilde x-y)-\zeta_A y,\\
	\frac d {dt} w &=& 
	\eta_B(1-\tilde x-w)-\zeta_B w.
\end{eqnarray*}
It is obvious that $y$ and $w$ tend to the stationary state, which is unique as $\tilde x$ is prescribed and fixed. If we define the lumped parameters  $\theta_A=\frac{\zeta_A}{\eta_A+\zeta_A}$, 
$\theta_B=\frac{\zeta_B}{\eta_B+\zeta_B}$, we have 
$$\lim_{t\rightarrow\infty} y(t) =  \theta_A\, \tilde x,
\qquad 
\lim_{t\rightarrow\infty} w(t) = 
\theta_B\,(1-\tilde x).$$
The slow manifold is given by $y =  \theta_A\, \tilde x$, and $ w(t)=
\theta_B\,(1-\tilde x).$  With the slow time $T=\eps t$, we find on the slow manifold for $\eps\rightarrow 0$ (recall $x=\tilde x-y=\theta_A\tilde x$, 
and $z=1-\tilde x-w=
\theta_B\,(1-\tilde x)$),
$$\frac d{dT} \tilde x = 
\mu\,\theta_A\theta_B \frac{
	g_A(\theta_A\, \tilde x, \theta_B\, (1-\tilde x))
	-
	g_B(\theta_A\, \tilde x, \theta_B\, (1-\tilde x)}
{1-\theta_A \tilde x-\theta_B (1-\tilde x)}\,\,\tilde x\,(1-\tilde x).
$$
A further state-dependent time transformation yields the modified replicator equation. We denote the time after that transformation again by $T$. 
\begin{cor}
	The modified replicator equation for the quiescence model is given by
	\begin{eqnarray}
		\frac d{dT} \tilde x = 
		[\,g_A(\theta_A\, \tilde x, \theta_B\, (1-\tilde x))
		-
		g_B(\theta_A\, \tilde x, \theta_B\, (1-\tilde x))\,]\,\,\,\tilde x\,(1-\tilde x).
	\end{eqnarray}
\end{cor}

The arguments of the replicator equation are thus only given by the active part of the population.  This result is quite intuitive. 
Last, we formulate explicitly the replicator equation for a given pay-off matrix, which we use in the next section to investigate specifically the case of cooperation.

\begin{theorem}
	Let  
	$$g_A(x,z) = (1,0)\,\left(\begin{array}{cc}
		A & B\\
		C & D
	\end{array}\right)\,\left(\begin{array}{c}
		x\\z\end{array}\right),\quad
	g_B(x,z) = (0,1)\,\left(\begin{array}{cc}
		A & B\\
C & D
	\end{array}\right)\,\left(\begin{array}{c}
		x\\z\end{array}\right). 
	$$
	Then we have the following modified replicator equation for the frequency of type~A individuals
\begin{eqnarray}
	\frac d{dT}\, \tilde x &=& 
[\,  \theta_A\,(A-C)\,\tilde x+\theta_B\,(B-D)\,(1-\tilde x)\,]\,\,\,\tilde x\,(1-\tilde x).
\end{eqnarray}
\end{theorem}

\subsection{Quiescence II -- active population is constant}
In general, quiescent individuals have a minimal need for resources. Therefore it might be more realistic that the population is bound only by its active part. In the second quiescence model we thus only keep the active population size constant, while the number of quiescent individuals can be arbitrary large. The transition probabilities for a stochastic version of the model are given in table~\ref{quiescIItrans2}.

\begin{table}[h!]
	\begin{adjustbox}{width=0.95\textwidth,center}
		\begin{tabular}{lll}
	\hline
	Type & Transition & Rate\\
	\hline
	death A \& birth type B 
	& $(X_t,Y_t,Z_t, W_t)\rightarrow(X_t-1,Y_t,Z_t+1, W_t)$
	&$\mu X_t\,\frac{\beta Z_t}{\beta(X_t+Z_t)+\zeta_AY_t+\zeta_BW_t}$\\
	death A \& wake-up type B 
    & $(X_t,Y_t,Z_t, W_t)\rightarrow(X_t-1,Y_t,Z_t+1, W_t-1)$
    &$\mu X_t\,\frac{\zeta_B W_t}{\beta(X_t+Z_t)+\zeta_AY_t+\zeta_BW_t}$\\
    	death A \& wake-up type A 
    & $(X_t,Y_t,Z_t, W_t)\rightarrow(X_t,Y_t-1,Z_t, W_t)$
    &$\mu X_t\,\frac{\zeta_AY_t}{\beta(X_t+Z_t)+\zeta_AY_t+\zeta_BW_t}$\\
    	death B \& birth type A 
    & $(X_t,Y_t,Z_t, W_t)\rightarrow(X_t+1,Y_t,Z_t-1, W_t)$
    &$\mu Z_t\,\frac{\beta X_t}{\beta(X_t+Z_t)+\zeta_AY_t+\zeta_BW_t}$\\
    death B \& wake-up type A 
    & $(X_t,Y_t,Z_t, W_t)\rightarrow(X_t+1,Y_t-1,Z_t-1, W_t)$
    &$\mu Z_t\,\frac{\zeta_AY_t}{\beta(X_t+Z_t)+\zeta_AY_t+\zeta_BW_t}$\\
    death B \& wake-up type B 
    & $(X_t,Y_t,Z_t, W_t)\rightarrow(X_t,Y_t,Z_t, W_t-1)$
    &$\mu Z_t\,\frac{\zeta_B W_t}{\beta(X_t+Z_t)+\zeta_AY_t+\zeta_BW_t}$\\
	go quiescent type A  \& birth type A
	& $(X_t,Y_t,Z_t, W_t)\rightarrow(X_t,Y_t+1,Z_t, W_t)$
	&$\eta_A X_t\, \frac{\beta X_t}{\beta(X_t+Z_t)+\zeta_AY_t+\zeta_BW_t}$\\
	go quiescent type A  \& birth type B 
& $(X_t,Y_t,Z_t, W_t)\rightarrow(X_t-1,Y_t+1,Z_t+1, W_t)$
&$\eta_A X_t\, \frac{\beta Z_t}{\beta(X_t+Z_t)+\zeta_AY_t+\zeta_BW_t}$\\
	go quiescent type A  \& wake-up type B 
& $(X_t,Y_t,Z_t, W_t)\rightarrow(X_t-1,Y_t+1,Z_t+1, W_t-1)$
&$\eta_A X_t\, \frac{\zeta_B W_t}{\beta(X_t+Z_t)+\zeta_AY_t+\zeta_BW_t}$\\
	go quiescent type B   \& birth type A 
	& $(X_t,Y_t,Z_t, W_t)\rightarrow(X_t+1,Y_t,Z_t-1, W_t+1)$
	&$\eta_B Z_t\, \frac{\beta X_t}{\beta(X_t+Z_t)+\zeta_AY_t+\zeta_BW_t}$\\
	go quiescent type B   \& birth type B 
& $(X_t,Y_t,Z_t, W_t)\rightarrow(X_t,Y_t,Z_t, W_t+1)$
&$\eta_B Z_t\, \frac{\beta Z_t}{\beta(X_t+Z_t)+\zeta_AY_t+\zeta_BW_t}$\\
	go quiescent type B   \& wake-up type A 
& $(X_t,Y_t,Z_t, W_t)\rightarrow(X_t+1,Y_t-1,Z_t-1, W_t+1)$
&$\eta_B Z_t \, \frac{\zeta_A Y_t}{\beta(X_t+W_t)+\zeta_AY_t+\zeta_BW_t}$
	\end{tabular}
\end{adjustbox}
\caption{Transitions for the quiescence model II.}\label{quiescIItrans2}
\end{table}

The corresponding ODE reads
\begin{eqnarray*}
x' &=& \frac{
- \mu x(\beta z+\zeta_B w)
+ \mu z(\beta x+\zeta_A y)
- \eta_A x(\beta z+\zeta_B w)
+ \eta_B z(\beta x+\zeta_A y)
}{\beta(x+z)+\zeta_A y+\zeta_B w},
\\
y' &=& 
\frac{
-\mu (z+x)\zeta_A y
+\eta_A x (\beta x+\beta z+\zeta_B w) 
-\eta_B z\zeta_A y
}{\beta(x+z)+\zeta_A y+\zeta_B w},
\\
z' &=& 
\frac{
	\mu x(\beta z+\zeta_B w)
	-\mu z(\beta x+\zeta_A y)
	+ \eta_A x(\beta z+\zeta_B w)
	- \eta_B z(\beta x+\zeta_A y)
}{\beta(x+z)+\zeta_A y+\zeta_B w},
\\
w' &=& 
\frac{
-\mu (z+x)\zeta_B w
-\eta_A x\zeta_B w
+\eta_B z (\beta x+\beta z+\zeta_A y) 
}{\beta(x+z)+\zeta_A y+\zeta_B w}.
\end{eqnarray*}
We observe that $x+z$ is constant. As we keep the active population constant, we normalize that population size to $1$, so that $x+z=1$, such that we only need to follow $x$ and replace $z$ by $1-x$:
\begin{eqnarray*}
x' &=& \frac{
	- x\,w\,\zeta_B(\mu\,+\eta_A)
	+(1-x)\,y\, \zeta_A( \mu+\eta_B) 
	+x\,(1-x)\,\beta (\eta_B-\eta_A)
	}{\beta+\zeta_A y+\zeta_B w},
	\\
y' &=& 
	\frac{
         - y\,\,\mu\zeta_A
         +x\,\,\beta \eta_A
         +x\, w\,\, \eta_A  \zeta_B 
         - (1-x) \,y \,\,\eta_B \zeta_A
	}{\beta+\zeta_A y+\zeta_B w},
	\\
w' &=& 
\frac{
-w\,\,\mu \zeta_B
+ (1-x) \,\,\beta  \eta_B
+(1-x)\,y\,\,\eta_B \zeta_A
- x \, w \,\,\eta_A \zeta_B 
}{\beta+\zeta_A y+\zeta_B w}.
\end{eqnarray*}
A simple, state-dependent time transformation yields a polynomial r.h.s.,
\begin{eqnarray}
	x' &=& 
		- x\,w\,\zeta_B(\mu\,+\eta_A)
		+(1-x)\,y\, \zeta_A( \mu+\eta_B) 
		+x\,(1-x)\,\beta (\eta_B-\eta_A),
	\\
	y' &=& 
		- y\,\,\mu\zeta_A
		+x\,\,\beta \eta_A
		+x\, w\,\, \eta_A  \zeta_B 
		- (1-x) \,y \,\,\eta_B \zeta_A,
	\\
	w' &=& 
		-w\,\,\mu \zeta_B
		+ (1-x) \,\,\beta  \eta_B
		+(1-x)\,y\,\,\eta_B \zeta_A
		- x \, w \,\,\eta_A \zeta_B .
\end{eqnarray}
\begin{prop} The model has a line of stationary points, given by 
	\begin{eqnarray}
	(x,y,w) =  \bigg(x, \,\,\,x\,\frac{\beta\, \eta_A}{\mu\,\zeta_A}, 
	\,\,\,(1-x)\, \frac{\beta\,\eta_B}{\mu\,\zeta_B}\bigg)
		\end{eqnarray}
	where $x\in[0,1]$.
\end{prop}
{\bf Proof: } We set $x'=y'=z'=0$, and determine the non-negative solutions from the 
resulting algebraic equations. We consider the equations $y'=0$ and $w'=0$ as a linear system in $y$ and $w$ (assuming $x$ to be a constant), and solve this system for $y$ and $w$, 
$$ A \left(\begin{array}{c}
	y\\w
\end{array}\right)\,\,
= -\vec b$$
with 
\begin{eqnarray*}
A = \left(\begin{array}{cc}
-\mu\zeta_A -(1-x)\eta_B\zeta_A       &x\,\eta_A\zeta_B \\
(1-x)\eta_B\zeta_A                     &-\mu\zeta_B -x\eta_A\zeta_B
\end{array}\right) 
\mbox{ and }
\vec b
=
\left(\begin{array}{c}
	x\eta_A\beta\\
	(1-x)\eta_B\beta
\end{array}\right)
.
\end{eqnarray*}
The determinant of $A$ reads 
$\mbox{det}(A)
=
\mu^2\zeta_A\zeta_B+x\eta_A\zeta_B\mu\zeta_A + (1-x)\eta_B\zeta_A\mu\zeta_B
$, 
s.t.
$$
A^{-1} = \frac 1 {\mbox{det}(A)}\,\left(\begin{array}{cc}
	 -\mu\zeta_B -x\eta_A\zeta_B      &-x\,\eta_A\zeta_B \\
	-(1-x)\eta_B\zeta_A     &-\mu\zeta_A -(1-x)\eta_B\zeta_A 
\end{array}\right).
$$
Therewith,
\begin{eqnarray*}
y&=&\frac{x[(\mu\zeta_B+x\eta_A\zeta_B)\eta_A\beta
	+
	\eta_A\zeta_B(1-x) \eta_B\beta]}
{\mu^2\zeta_A\zeta_B+x\eta_A\zeta_B\mu\zeta_A + (1-x)\eta_B\zeta_A\mu\zeta_B}
= x\, \frac{\beta\eta_A }{\mu\zeta_A}\\
w&=&\frac{(1-x)[x\eta_A\beta\eta_B\zeta_A 
+
(\mu\zeta_A +(1-x)\eta_B\zeta_A)\eta_B\beta]}
{\mu^2\zeta_A\zeta_B+x\eta_A\zeta_B\mu\zeta_A + (1-x)\eta_B\zeta_A\mu\zeta_B}
= (1-x)\, \frac{\beta\eta_B }{\mu\zeta_B}.
\end{eqnarray*}
We plug that result into the r.h.s.\ of the equation for $x'$, and obtain
\begin{eqnarray*}
	&&	- x\,w\,\zeta_B(\mu\,+\eta_A)
	+(1-x)\,y\, \zeta_A( \mu+\eta_B) 
	+x\,(1-x)\,\beta (\eta_B-\eta_A)\\
&=& x(1-x)\,\{
- 
\zeta_B(\mu\,+\eta_A)\, \frac{\beta\eta_B }{\mu\zeta_B}
+
\zeta_A( \mu+\eta_B)\ \frac{\beta\eta_A }{\mu\zeta_A}
 +\beta (\eta_B-\eta_A)
\} = 0.
\end{eqnarray*}
That is, with the choice for $y$ and $z$, we always have $x'=0$. Therefore, we always have a line of stationary points.
\par\qed\par\medskip 

This proposition seems to indicate that quiescence models tend to couple the active and resting populations more tightly than seed bank models (see below).

\subsubsection{Weak selection}

We introduce weak selection in the birth term. Let $g_A=g_A(x,z)$ resp.\ $g_B=g_B(x,z)$ indicate the frequency-dependent selection terms. We then replace $\beta x$ by $(\beta+\eps g_A)x$ and $\beta z$ by $(\beta+\eps g_B)z$. Let $$F = (\beta+\eps g_A) x+(\beta+\eps g_B) z+\zeta_A y+\zeta_B w.$$
Therewith, the starting point for our model becomes (we multiply the equations by $F$, that is, by the denominator of the r.h.s.)
\begin{eqnarray*}
	F\,\cdot\,x' &=& 
		- \mu x((\beta+\eps g_B) z+\zeta_B w)
		+ \mu z((\beta+\eps g_A) x+\zeta_A y)\\
&&		- \eta_A x((\beta+\eps g_B) z+\zeta_B w)
		+ \eta_B z((\beta+\eps g_A) x+\zeta_A y),
	\\
	F\,\cdot\,y' &=& 
		-\mu (z+x)\zeta_A y
		+\eta_A x ((\beta+\eps g_A) x+(\beta+\eps g_B) z+\zeta_B w) 
		-\eta_B z\zeta_A y,
	\\
F\,\cdot\,z' &=& 
		\mu x(\beta z+\zeta_B w)
		-\mu z((\beta+\eps g_A) x+\zeta_A y)\\
&&		+ \eta_A x((\beta+\eps g_B) z+\zeta_B w)
		- \eta_B z((\beta+\eps g_A) x+\zeta_A y),
	\\
F\,\cdot\,	w' &=& 
		-\mu (z+x)\zeta_B w
		-\eta_A x\zeta_B w
		+\eta_B z ((\beta+\eps g_A) x+(\beta+\eps g_B) z+\zeta_A y) .
\end{eqnarray*}
We again use that $x+z=1$, and simplify the equation by a state-dependent 
time transformation that removes the factor $F$. Therewith, the model with weak selection becomes (herein, we have $g_A= g_A(x,1-x))$, $g_B= g_B(x,1-x))$)
\begin{eqnarray}
	x' &=& 
		- x\,w\,\zeta_B(\mu\,+\eta_A)
	+(1-x)\,y\, \zeta_A( \mu+\eta_B) 
	+x\,(1-x)\,\beta (\eta_B-\eta_A)\\
&& + \eps\,\,\,x(1-x)\, ( g_A\, (\mu+\eta_B)- g_B\,(\mu+\eta_A)),\nonumber 
	\\
y' &=& 
		- y\,\,\mu\zeta_A
+x\,\,\beta \eta_A
+x\, w\,\, \eta_A  \zeta_B 
- (1-x) \,y \,\,\eta_B \zeta_A  \\
&&+\eps \,\,\,\eta_A   \, ( g_A\, x+ g_B\,(1-x))\,x,  \nonumber\\
	w' &=& 
		-w\,\,\mu \zeta_B
+ (1-x) \,\,\beta  \eta_B
+(1-x)\,y\,\,\eta_B \zeta_A
- x \, w \,\,\eta_A \zeta_B \\
&&
+ \eps\,\,\, 	\eta_B (g_A x+g_B (1-x))\, (1-x).\nonumber 
\end{eqnarray}

We introduce new variables to separate processes at different time scales,
$$ x = \xt,\quad 
y = x\,\frac{\beta\, \eta_A}{\mu\,\zeta_A}+\eps \yt,\quad 
w = (1-x)\,\frac{\beta\,\eta_B}{\mu\,\zeta_B}+\eps \wt.$$

\begin{prop} \label{quiesceceIItrafo}The dynamical system in the new variables reads
\begin{eqnarray}
\xt' &=& 
\eps \bigg(
- \xt\,\wt\,\zeta_B(\mu\,+\eta_A)
+(1-\xt)\,\yt\, \zeta_A( \mu+\eta_B)
\\
&&\qquad\quad +\xt(1-\xt)\, ( g_A (\mu+\eta_B)- g_B(\mu+\eta_A))\bigg),\nonumber 
\\
\yt' &=& 
-\mu\zeta_A\, \yt
+
\eta_A\zeta_B\,
\frac{\beta(\mu+\eta_A)+\mu\zeta_A}{\mu\zeta_A}
\xt\,\wt
- \frac{\beta\eta_A(\mu+\eta_B)+\mu\eta_B\zeta_A }{\mu} (1-\xt)\,\yt\\
&&  +
\eta_A   \, ( g_A\, \xt+ g_B\,(1-\xt))\,\xt
-\frac{\beta\, \eta_A}{\mu\,\zeta_A}
\xt(1-\xt)\, ( g_A (\mu+\eta_B)- g_B(\mu+\eta_A)),\nonumber
\\
\wt' &=&	
-\mu\zeta_B\, \wt 
+\eta_B\zeta_A \frac{\beta(\mu+\eta_B)+\mu\zeta_B }{\mu\zeta_B} (1-\xt)\,\yt
-
\frac{\beta\eta_B(\mu+\eta_A)+\mu\eta_A\zeta_B}
{\mu}
\xt\,\wt\\
&&+
\eta_B   \, ( g_A\, \xt+ g_B\,(1-\xt))\,(1-\xt)
+\frac{\beta\, \eta_B}{\mu\,\zeta_B}
\xt(1-\xt)\, ( g_A (\mu+\eta_B)- g_B(\mu+\eta_A)).\nonumber
\end{eqnarray}
\end{prop}
The proof of this proposition is fairly straightforward but lengthy, and can be found in the appendix.

\begin{prop}
The slow manifold is transversally stable in the fast system  and given by $\yt=Y(\xt)$, $\wt=W(\xt)$, which are defined as the unique solution of 
$$ A \left(\begin{array}{c} Y(\xt)\\W(\xt)
	\end{array}\right)= -\vec b(\xt),$$
where 
$$ A = \left(\begin{array}{cc}
	- \frac{\beta\eta_A(\mu+\eta_B)+\mu\eta_B\zeta_A }{\mu} (1-\xt)-\mu\zeta_A & \eta_A\zeta_B\,
	\frac{\beta(\mu+\eta_A)+\mu\zeta_A}{\mu\zeta_A}
	\xt\\
	\eta_B\zeta_A \frac{\beta(\mu+\eta_B)+\mu\zeta_B }{\mu\zeta_B} (1-\xt) & -
	\frac{\beta\eta_B(\mu+\eta_A)+\mu\eta_A\zeta_B}
	{\mu}
	\xt -\mu\zeta_B
\end{array}\right),$$
and
$$
\vec b(\xt) = \left(\begin{array}{c} \eta_A   \, ( g_A\, \xt+ g_B\,(1-\xt))\,\xt
	-\frac{\beta\, \eta_A}{\mu\,\zeta_A}
	\xt(1-\xt)\, ( g_A (\mu+\eta_B)- g_B(\mu+\eta_A))\\ 
	\eta_B   \, ( g_A\, \xt+ g_B\,(1-\xt))\,(1-\xt)
	+\frac{\beta\, \eta_B}{\mu\,\zeta_B}
	\xt(1-\xt)\, ( g_A (\mu+\eta_B)- g_B(\mu+\eta_A))
\end{array}\right).$$
\end{prop}
{\bf Proof:} If we take $\eps$ to zero, then $\xt$ is frozen. We determine the limit of $\yt$ and $\zt$. They satisfy a linear equation, 
$$ \frac d {dt}\left(\begin{array}{c} \tilde y\\ \tilde z\end{array}\right) 
= 
A \left(\begin{array}{c} \tilde y\\ \tilde z\end{array}\right)+\vec b(\xt)$$
where $A$ and $\vec b(\xt)$ is given above. 
We find
\begin{eqnarray*}
	\mbox{det}(A)
	&=& \mu^2\zeta_A\zeta_B 
+\zeta_B\,(\beta\eta_A\,(\mu+\eta_B)+\mu\eta_B\zeta_A)\,(1-\xt)\\
&&+\zeta_A\,(\beta\eta_B\,(\mu+\eta_A)+\mu\eta_A\zeta_B)\,\xt
+ \beta\,(\eta_A-\eta_B)\,(\eta_A\,\zeta_B-\eta_B\zeta_A)\,\xt(1-\xt).
\end{eqnarray*}
We can check that $\mbox{det}(A)>0$ by taking $\mu=0$ (which only decreases the determinant), and multiply out the remaining terms. Furthermore, $\mbox{tr}(A)<0$, so that the solution of this system tends to its unique stationary point $-A^{-1}\vec b(\xt)$. 

\par\qed\par\medskip 

\begin{theorem}
The dynamics of the slow manifold is given by 
\begin{eqnarray}
\frac d {dT} \xt &=& 
(g_A(\xt,1-\xt)-g_B(\xt,1-\xt))\,\xt(1-\xt).
\end{eqnarray}
\end{theorem}
The proof is performed by computer algebra (MAXIMA~\cite{maxima}),
where we obtain 
\begin{eqnarray*}
\frac d {dT} \xt 
&=& \frac{\mu\zeta_A\zeta_B\,(\mu+\eta_A\xt+\eta_B\,(1-\xt))}
{\mu\zeta_A\zeta_B+\beta\,(\eta_A\zeta_B\,(1-\xt)+\eta_B\,\zeta_A\xt)}\,\,
(g_A(\xt,1-\xt)-g_B(\xt,1-\xt))\,\xt(1-\xt).
\end{eqnarray*}
A state-dependent time transformation yields the result. In other words, the replicator equation is not affected by the life-history trait. Clearly, for the well-known payoff matrix, we find back the usual replicator equation.

\begin{theorem}
	Let  
	$$g_A(x) = (1,0)\,\left(\begin{array}{cc}
		A & B\\
		C & D
	\end{array}\right)\,\left(\begin{array}{c}
		x\\1-x\end{array}\right),\quad
	g_B(x) = (0,1)\,\left(\begin{array}{cc}
		A & B\\
		C & D
	\end{array}\right)\,\left(\begin{array}{c}
		x\\1-x\end{array}\right). 
	$$
	Then we have the following modified replicator equation
	\begin{eqnarray}
		\frac d{dT}\, \tilde x &=& 
		[\,  (A-C)\,x+\,(B-D)\,(1-x)\,]\,\,\,\tilde x\,(1-\tilde x).
	\end{eqnarray}
\end{theorem}

\subsection{Seed bank}
Now we turn to a plant population with a seed bank. 
We assume that the above-ground population has fixed size $N$. We have $X_t$ type-A above-ground individuals, and $N-X_t$  type-B above-ground individuals; below-ground, the number of type-A seeds is $Y_t$, and that of type-B seeds $W_t$. Above-ground individuals die at rate $\mu$, and any seed has the same chance to germinate and to replace a dead above-ground individual. Seeds are produced at rate $\beta_A$ (type A) resp.\ $\beta_B$ (type B), and die at rate $\mu_A$ (type A), resp.\ $\mu_B$ (type B). 
For a stochastic model, we find the following transitions  \par\medskip 

\begin{table}[h!]
	\begin{adjustbox}{width=0.95\textwidth,center}
		\begin{tabular}{lll}
\hline
Type & Transition & Rate\\
\hline
Death A \& Birth type A 
& $(X_t,Y_t,W_t)\rightarrow(X_t,Y_t-1,W_t)$
&$\mu X_t\,\frac{Y_t}{Y_t+W_t}$\\
Death B \& Birth type B
& $(X_t,Y_t,W_t)\rightarrow(X_t,Y_t,W_t-1)$
&$\mu (N-X_t)\,\frac{W_t}{Y_t+W_t}$\\
Death A \& Birth type B
& $(X_t,Y_t,W_t)\rightarrow(X_t-1,Y_t,W_t-1)$
&$\mu X_t\,\frac{W_t}{Y_t+W_t}$\\
Death B \& Birth type A
& $(X_t,Y_t,W_t)\rightarrow(X_t+1,Y_t-1,W_t)$
&$\mu (N-X_t)\,\frac{Y_t}{Y_t+W_t}$\\
Seed product.\ type A
& $(X_t,Y_t,W_t)\rightarrow(X_t,Y_t+1,W_t)$
&$\beta_A X_t$\\
Seed product.\ type B
& $(X_t,Y_t,W_t)\rightarrow(X_t,Y_t,W_t+1)$
&$\beta_B (N-X_t)$\\
Seed death.\ type A
& $(X_t,Y_t,W_t)\rightarrow(X_t,Y_t-1,W_t)$
&$\mu_A Y_t$\\
Seed death.\ type B
& $(X_t,Y_t,W_t)\rightarrow(X_t,Y_t,W_t-1)$
&$\mu_B W_t$\\
\hline 
\end{tabular}
\end{adjustbox}
\caption{Transitions for the seed bank model.}\label{seedbank}
\end{table}

\par\medskip 

Note that we encounter an issue if the model jumps into a state with $Y_t=W_t=0$. However, if $N$ is large enough (particularly if we consider the transition to the deterministic limit), the probability for a transition into that state becomes arbitrary small and tends to zero. The corresponding deterministic model reads
\begin{eqnarray*}
\frac d {dt} X &=& 
-\mu X\,\frac{Z}{Y+Z}+\mu\,(N-X)\,\frac{Y}{Y+Z},\\
\frac d {dt} Y &=& 
- \mu X\,\frac{Y}{Y+Z}\,-\, \mu\,(N-X)\,\frac{Y}{Y+Z}+ \beta_AX-\mu_AY,\\
\frac d {dt} Z &=& 
-\, \mu\,(N-X)\,\frac{Z}{Y+Z}\,-\, \mu X\,\frac{Z}{Y+Z}\,+ \beta_B(N-X)-\mu_BZ.
\end{eqnarray*}
We rescale the variables, $x(t)=X(t)/N$,  $y(t)=Y(t)/N$, 
and  $z(t)=Z(t)/N$ and obtain 
\begin{eqnarray}
	\frac d {dt} x &=& 
	-\mu x + \mu\,\frac{y}{y+z},\\
	\frac d {dt} y &=& 
	-\, \mu\,\frac{y}{y+z}+ \beta_Ax-\mu_Ay,
	\\
	\frac d {dt} z &=& 
	-\, \mu\,\frac{z}{y+z}\,+ \beta_B(1-x)-\mu_Bz.
\end{eqnarray}
The analysis of this model is slightly more involving than that for the quiescence models above. We start off with the stationary points of the system.

\begin{prop} We always have the one-species stationary points $(x,y,z)=(1,\frac{\mu+\mu_A}{\beta_A},0)$ resp.
	 $(x,y,z)=(0,0, \frac{\mu+\mu_B}{\beta_B})$. Only in the ungeneric situation 
	 $$\frac{\beta_A-\mu}{\mu_A} = \frac{\beta_B-\mu}{\mu_B}$$
	 we find a line of stationary points. Defining $\theta := \frac{\beta_A-\mu}{\mu_A}= \frac{\beta_B-\mu}{\mu_B}$, 
	$$ (x,y,z) = (x,\,\theta x,\, \theta(1-x))$$
	are stationary points  for all $x\in[0,1]$.
\end{prop}
{\bf Proof:} 
From $\dot x=0$ we obtain 
$	\mu x = \mu\,\frac{y}{y+z}$
and from $\dot y=0$ we have 
$ -\mu x+ \beta_Ax-\mu_Ay.$ Thus,
$y = \frac{\beta_A-\mu}{\mu_A} x$.
Similarly, from $\dot z = 0$ we get $ -\mu (1-x)+ \beta_B(1-x)-\mu_Bz$ s.t.\ 
$z = \frac{\beta_B-\mu}{\mu_B} (1-x)$. 
Altogether, we conclude 
$$ \mu x = 
\frac
{\frac{\beta_A-\mu}{\mu_A} x}
{\frac{\beta_A-\mu}{\mu_A} x +\frac{\beta_B-\mu}{\mu_B} (1-x) }
$$
Clearly, $x=0$ and $x=1$ always solve this equation. Only in the case that 
$\frac{\beta_A-\mu}{\mu_A} = \frac{\beta_B-\mu}{\mu_B}$ 
a whole line of stationary points is obtained, as the equation is true for all $x\in[0,1]$. \par\qed\par\medskip

\subsubsection{Weak selection}
We add now weak frequency-dependent selection. As before, we incorporate the selection in the reproduction process, so that the reproduction rates $\beta_A$ and $\beta_B$ are weakly modified by the pay-off functions $g_A(x)$ and $g_B(x)$. Note that these functions only depend on the above-ground population that is completely characterized by the fraction $x$ of type-A individuals. We obtain the model
\begin{eqnarray}
	\frac d {dt} x &=& 
	-\mu x + \mu\,\frac{y}{y+z},\\
	\frac d {dt} y &=& 
	-\, \mu\,\frac{y}{y+z}+ (\beta_A+ \eps\,g_A(x))x-\mu_Ay,
	\\
	\frac d {dt} z &=& 
	-\, \mu\,\frac{z}{y+z}\,+ (\beta_B+\eps\,g_B(x))(1-x)-\mu_Bz.
\end{eqnarray}
As before, we aim to apply singular perturbation theory. Thereto we clearly separate slow and fast variables, using the coordinate transformation 
$$
 x(t)= \tilde x(t),\qquad 
y(t)=\theta \tilde x(t)+\eps\,\tilde y(t),\qquad 
z(t)=\theta (1-\tilde x(t))+\eps\,\tilde z(t).
$$
\begin{prop}\label{seedTrafo} Assuming $\theta := \frac{\beta_A-\mu}{\mu_A}= \frac{\beta_B-\mu}{\mu_B}$, then we obtain, 
\begin{eqnarray}
	\frac d {dt} \tilde x &=& 
	 \mu\,\eps\,\frac{\tilde y\,(1-\tilde x)\,-\,\tilde z\,\tilde x}{\theta +\eps\,(\tilde y+\tilde z)},
	\\
	\frac d {dt} \tilde y &=& 
	-\, \,\mu\,(1+\theta)\,\frac{\tilde y(1-\tilde  x)- \tilde x\,\tilde z}
	{\theta +\eps\,(\tilde y+\tilde z)}
	+ \,g_A(\tilde x)\tilde x-\mu_A \,\tilde y,
	\\
	\frac d {dt} \tilde z &=& 
	 \, \mu\,(1+\theta)\,\frac{\tilde y\,(1-\tilde x)\,-\,\tilde z\,\tilde x}{\theta +\eps\,(\tilde y+\tilde z)}\, 
	+ g_B(x)(1-\tilde x)-\mu_B\,\tilde z .
\end{eqnarray}
\end{prop}
The proof of this proposition can be found in the appendix. We then first analyse the fast system. If we take the limit $\eps\rightarrow 0$, we obtain
\begin{eqnarray}
	\frac d {dt} \tilde x &=& 0,
	\\
	\frac d {dt} \tilde y &=& 
	-\, \,\mu\,\frac{(1+\theta)}{\theta}\,(\tilde y(1-\tilde  x)- \tilde x\,\tilde z)
	+ \,g_A(\tilde x)\tilde x-\mu_A \,\tilde y,
	\\
	\frac d {dt} \tilde z &=& 
	\, \mu\,\frac{(1+\theta)}{\theta}\,(\tilde y\,(1-\tilde x)\,-\,\tilde z\,\tilde x)\, 
	+ g_B(x)(1-\tilde x)-\mu_B\,\tilde z .
\end{eqnarray}
\begin{prop}
The slow manifold is given by 
\begin{eqnarray}
\tilde y &=& 
\frac{
	\mu\, \frac{(1+\theta)}{\theta}\,
	( g_A(\tilde x)\tilde x + g_B(\tilde x)(1-\tilde x) )
	+\mu_B  g_A(\tilde x)}
{\mu_A\,\mu_B\, 
	+\mu\, \frac{(1+\theta)}{\theta}\, (\mu_A \tilde x+\mu_B (1-\tilde x))}\,\,\tilde x, \\
\tilde z &=& 
\frac{\mu\, \frac{(1+\theta)}{\theta}\,
	( g_A(\tilde x)\tilde x + g_B(\tilde x)(1-\tilde x) )
	+\mu_A  g_B(\tilde x)}
{\mu_A\,\mu_B\, 
	+\mu\, \frac{(1+\theta)}{\theta}\, (\mu_A \tilde x+\mu_B (1-\tilde x))}\,\,(1-\tilde x).
\end{eqnarray}
\end{prop}
{\bf Proof:} 
As $\tilde x$ is constant in the fast system, the ODE becomes a linear system in $\tilde y$, $\tilde z$, which can be written as
$$ \frac d {dt}\left(\begin{array}{c} \tilde y\\ \tilde z\end{array}\right) 
= 
A \left(\begin{array}{c} \tilde y\\ \tilde z\end{array}\right)+b,$$
where
$$ A = \left(\begin{array}{cc}
-\,\mu\,\frac{(1+\theta)}{\theta}(1-\tilde x)-\mu_A & \mu\,\frac{(1+\theta)}{\theta}\tilde x\\
\mu\,\frac{(1+\theta)}{\theta}(1-\tilde x) & -\mu\,\frac{(1+\theta)}{\theta}\tilde x -\mu_B
\end{array}\right),\qquad 
b = \left(\begin{array}{c} g_A(\tilde x)\tilde x\\ 
	g_B(\tilde x)(1-\tilde x)\end{array}\right).$$
Since $\mbox{tr}(A)<0$ and $\mbox{det}(A)= \mu_A\,\mu_B\, 
+\mu\, \frac{(1+\theta)}{\theta}\, (\mu_A \tilde x+\mu_B (1-\tilde x))>0$, the unique stationary point is globally stable. The slow manifold is thus given by 
$$ 
\left(\begin{array}{c} \tilde y\\ \tilde z\end{array}\right) = -A^{-1} 
\left(\begin{array}{c} g_A(\tilde x)\tilde x\\ 
	g_B(\tilde x)(1-\tilde x)\end{array}\right)
,
$$
with 
$$ A^{-1}
=
\frac{1}{\det(A)}
 \left(\begin{array}{cc}
	-\mu\,\frac{(1+\theta)}{\theta}\tilde x -\mu_B
	& -\mu\,\frac{(1+\theta)}{\theta}\tilde x\\
	-	\mu\,\frac{(1+\theta)}{\theta}(1-\tilde x) & 
	-\,\mu\,\frac{(1+\theta)}{\theta}(1-\tilde x)-\mu_A
\end{array}\right),
$$
and $\mbox{det}(A)$ as stated above. We obtain the desired result.
\par\qed\par\medskip

We now turn to the slow system on the slow manifold. 
\begin{prop} Let $t=\eps T$. The slow system is given by 
\begin{eqnarray}
\frac d {dT} \tilde x &=& \mu 
\frac{
	\mu_B g_A(\tilde x)-\mu_A g_B(\tilde x)
}
{\theta \mu_A\,\mu_B\, 
	+\mu\, (1+\theta)\, (\mu_A \tilde x+\mu_B (1-\tilde x))}\,
\tilde x\,(1-\tilde x).
\end{eqnarray}
\end{prop}
{\bf Proof: } We investigate $\frac d {dt} \tilde x = 
\mu\,\eps\,\frac{\tilde y\,(1-\tilde x)\,-\,\tilde z\,\tilde x}{\theta +\eps\,(\tilde y+\tilde z)}$ 
on the slow manifold. On this manifold, we observe that 
\begin{eqnarray*}
	&& \tilde y (1-\tilde x)-\tilde z \tilde x\\
&=&\bigg(
\frac{
	\mu\, \frac{(1+\theta)}{\theta}\,
	( g_A(\tilde x)\tilde x + g_B(\tilde x)(1-\tilde x) )
	+\mu_B  g_A(\tilde x)}
{\mu_A\,\mu_B\, 
	+\mu\, \frac{(1+\theta)}{\theta}\, (\mu_A \tilde x+\mu_B (1-\tilde x))}\\
&&\qquad\qquad\qquad\qquad\qquad-
\frac{\mu\, \frac{(1+\theta)}{\theta}\,
	( g_A(\tilde x)\tilde x + g_B(\tilde x)(1-\tilde x) )
	+\mu_A  g_B(\tilde x)}
{\mu_A\,\mu_B\, 
	+\mu\, \frac{(1+\theta)}{\theta}\, (\mu_A \tilde x+\mu_B (1-\tilde x))}
\bigg)\,\tilde x\,(1-\tilde x)\\
&=& 
\frac{
\mu_B g_A(\tilde x)-\mu_A g_B(\tilde x)
}
{\mu_A\,\mu_B\, 
	+\mu\, \frac{(1+\theta)}{\theta}\, (\mu_A \tilde x+\mu_B (1-\tilde x))}\,
\tilde x\,(1-\tilde x). 
\end{eqnarray*} 
With the time transformation $T=\eps t$ and $\eps\rightarrow 0$, we find 
$$ 	\frac d {dT} \tilde x =
\mu\,\frac{\tilde y\,(1-\tilde x)\,-\,\tilde z\,\tilde x}{\theta}.$$
Hence, the result follows.
\par\qed\par\medskip

Note that a further (state-dependent) time transformation is able to remove the denominator.
\begin{theorem} The slow equation is given by 
	\begin{eqnarray}
		\frac d {dT} \tilde x &=& 
	[\,\mu_B\, g_A(\tilde x)-\mu_A\, g_B(\tilde x)\,]
	\,\,\xt\,(1-\xt).
	\end{eqnarray}
\end{theorem}

We formulate the replicator equation explicitly for a given pay-off matrix.

\begin{theorem}
	Let  
	$$g_A(x) = (1,0)\,\left(\begin{array}{cc}
		A & B\\
		C & D
	\end{array}\right)\,\left(\begin{array}{c}
		x\\1-x\end{array}\right),\quad
	g_B(x) = (0,1)\,\left(\begin{array}{cc}
		A & B\\
		C & D
	\end{array}\right)\,\left(\begin{array}{c}
		x\\1-x\end{array}\right). 
	$$
	Then we have the following modified replicator equation
	\begin{eqnarray}
		\frac d{dT}\, \tilde x &=& 
		[\,  (\mu_B\,A-\mu_A\,C)\,x+\,(\mu_B B-\mu_A D)\,(1-x)\,]\,\,\,\tilde x\,(1-\tilde x).
	\end{eqnarray}
\end{theorem}

\section{Evolution of cooperation}
We use the standard prisoners dilemma model to investigate cooperation as a trait under frequency-dependent selection~\cite{Hoesel2019}. Type A is the cooperator producing a public good at rate $b$ at cost $c$, and type B is the defector. The pay-off matrix is thus given by 
$$ 
\left(\begin{array}{cc}
	A & B\\
	C & D
\end{array}\right)
=
\left(\begin{array}{cc}
	b-c & -c\\
	b & 0
\end{array}\right).$$

We start the analysis of first quiescence model. 
\begin{theorem}
Under the quiescence model I, the replicator equation reads
\begin{eqnarray}
\frac d {dT}\,\tilde x = 
-c\,[\theta_A\, \tilde x+\theta_B\,(1-\tilde x)]\,\tilde x\,(1-\tilde x)
\end{eqnarray}
while for the quiescence model II, we find 
\begin{eqnarray}
	\frac d {dT}\,\tilde x = 
	-c\,\,\tilde x\,(1-\tilde x).
\end{eqnarray}
In other words, the cooperator always dies out, independently of the time scale.
\end{theorem}

\begin{figure}[htb]
\begin{center}
\includegraphics[width=8cm]{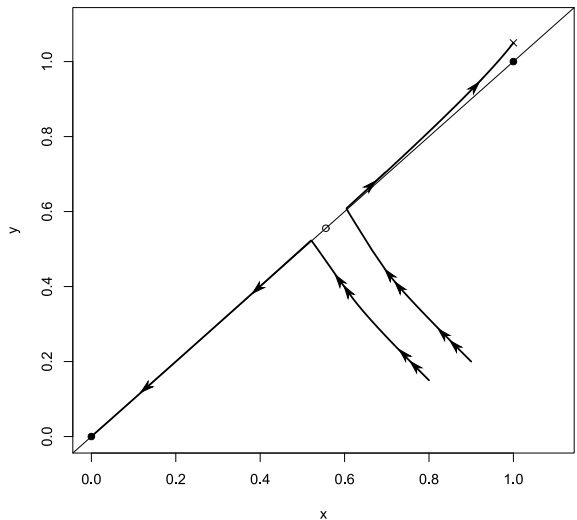}
\end{center}
\caption{Simulation of the seed bank model ($x$ and $y$ component). Bold lines: two trajectories, thin line: slow manifold. Closed circles: Locally stable stationary points on the slow manifold, open circle: unstable stationary point on the slow manifold (approximative equation). Cross: Stationary point of the original system. Parameters: 
$\mu = 1$, 
$\mu_A = 0.1$, 
$\mu_B = 1$, 
$\theta = 1$, 
$b =   2$, 
$c = 1.0$, 
$\varepsilon=0.005$.}\label{seedbFig}
\end{figure}

The picture changes for the seed bank model.
\begin{theorem}\label{seedCoop}
Under the seed bank model, the replicator equation reads 
\begin{eqnarray}
\frac d {dT}\,\tilde x = 
[(\mu_B\, (b-c)-\mu_A\,b)\, \tilde x - \mu_B\,c\,(1-\tilde x) ]\,\,\tilde x\,(1-\tilde x).
\end{eqnarray}
The defector-only solution $\xt=0$ is always locally asymptotically stable. The cooperator-only solution $\xt=1$ is either locally asymptotically stable or unstable. If 
 $$(\mu_B-\mu_A)\,b < \mu_B\, c,$$
the cooperator dies out and $\xt =0$ is globally stable in $[0,1)$. Under the reverse condition, 
 $$(\mu_B-\mu_A)\,b > \mu_B\, c,$$
 $\xt=0$ and $\xt=1$ both are locally asymptotically stable stationary states,  and there is an additional, unstable stationary point given by
$$ \tilde x = \frac{\mu_B\, c}{(\mu_B-\mu_A)\,b}.$$
\end{theorem}
In Fig.~\ref{seedbFig}, we see a simulation of the original system, together with the slow manifold. The parameters are chosen in such a way that the point $\xt=1$ is locally asymptotically stable, such that two locally stable and one unstable stationary points are located on the slow manifold. As expected, we find that the trajectories saddle down on the slow manifold on a fast time scale, and then slowly proceed, in the vicinity of the slow manifold, towards the stationary points. While in the left part of the slow manifold, the agreement of trajectory and slow manifold is perfect, in the right part we find slight residuals (deviations). The stationary point on the slow manifold is $(x,y,z)=(1,1,0)$ but the point approached by the simulation is estimated to be $(1,1.05,0)$ (meaning a difference of ${\cal O}(\varepsilon)$).

\subsection{Adaptive Dynamics}
We further refine our investigation. Instead of considering defectors and cooperators, we consider traits with different degree of cooperation. The degree of cooperation can be parametrized by the public good production $b$. The costs $C$ are naturally a strictly increasing, smooth function of $b$, $c=C(b)$ with $C(0)=0$. We assume concomitantly that also the time scale of quiescence/dormancy is a function of $b$. This assumption is less natural, but requires in biological terms some form of pleiotropy, namely that there is a set of genes determining both the quiescence/dormancy and the cooperation phenotypes. The pleiotropic assumption allows the co-evolution of both traits (cooperation and quiescence/dormancy) to occur (see discussion for a biological view of this assumption). We thereafter use adaptive dynamics to study the evolution of both traits. Type A is the resident (with abundance $\xt$) and has parameter $b_r$, while type B (with abundance $1-\xt$) plays the role of the mutant, and has parameter $b_m$. Therewith, the payoff-matrix is given by 
$$ 
\left(\begin{array}{cc}
	A & B\\
	C & D
\end{array}\right)
=
\left(\begin{array}{cc}
	b_r-C(b_r) & b_m-C(b_r)\\
	b_r-C(b_m) & b_m-C(b_m)
\end{array}\right).$$

\subsubsection{Quiescence models}

Under the quiescence model I, $\theta_A$ and $\theta_B$ express the time scales for the quiescent state in the replicator equation. 
We assume the time scales to be a function of $b$, such that $\theta_A=\theta(b_r)$ and $\theta_B=\theta(b_m)$ for some suited, strictly positive function $\theta:\R_+\rightarrow\R_+$. 
\begin{theorem}
In case of the quiescence model I and II, the strategy $b=0$ is a convergent stable ESS, which attracts all solutions.
\end{theorem}
{\bf Proof: }
The replicator equation for the quiescence model I reads 
$$
	\frac d{dT}\, \tilde x = (C(b_m)-C(b_r))\,[\,  \theta(b_r)\,\,\tilde x+\theta(b_m)\,\,(1-\tilde x)\,]\,\,\,\tilde x\,(1-\tilde x).
$$
while we obtain for quiescence model II
$$
\frac d{dT}\, \tilde x = (C(b_m)-C(b_r))\,\,\tilde x\,(1-\tilde x).
$$
The mutants can invade and take over if and only if they have smaller costs, so that the system always tends to the state with $b=0$. The choice of the time scale function $\theta(b)$ does not affect the result.
\par\qed\par\medskip

\subsubsection{Seed banks}

We assume that the time scale of the seed bank depends on $b$, namely $\mu_A=\mu_s(b_r)$, $\mu_B=\mu_s(b_m)$ for some smooth function $\mu_s$. As $\beta_{*}$ and $\mu_{*}$  for $*\in\{A,B\}$ are coupled via $(\mu+\beta_*)/\beta_* = \theta$, we find for both traits 
$$ \beta(b) = \mu_s(b)\,\theta-\mu.$$
Therewith, the replicator equation becomes 
\begin{eqnarray*}
	\frac d{dT} \xt
	&=&
	\bigg\{[\mu_s(b_r)\,(b_r-C(b_r)) - \mu_s(b_m)\,(b_r-C(b_m))]\,\xt \\
	&&\qquad + [\mu_s(b_r)\,( b_m-C(b_r)) - \mu_s(b_m)\,(b_m-C(b_m))]\,(1-\xt) \bigg\}
	\,\xt\,(1-\xt)\\
  &=& 	{\cal G}(\xt;b_r,b_m)\,\,\xt(1-\xt),
\end{eqnarray*}
with 
\begin{eqnarray*}
	{\cal G}(\xt;b_r,b_m)
	&=& 
	(\mu_s(b_r)\, - \mu_s(b_m))\,(b_m + (b_r-b_m)\,\xt) 
	+ [ \mu_s(b_m)\,C(b_m)  - \mu_s(b_r)\,C(b_r)].
\end{eqnarray*} 
Then we find 
\begin{eqnarray*}
\frac{\partial}{\partial b_m}	{\cal G}(\xt;b_r,b_m)\bigg|_{b_m=b_r}
	&=& 
	(C(b_r)-b_r)\,\mu_s'(b_r)\,  
	+ C'(b_r) \mu_s(b_r)
\end{eqnarray*} 
and 
$${\mathcal G}(\xt;b_r,b_m)
= (b_m-b_r)\,\frac{\partial}{\partial b_m}{\mathcal G}(\xt;b_r,b_m)\bigg|_{b_m=b_r}+{\cal O}((b_m-b_r)^2).$$
Note that $\frac{\partial}{\partial b_m}{\mathcal G}(\xt;b_r,b_m)\bigg|_{b_m=b_r}$ does not depend on $\xt$. That is, for $b_m$ sufficiently close to $b_r$, ${\mathcal G}(\xt;b_r,b_m)$ 
does not change the sign in $\xt\in[0,1]$ (this interval is compact). As a mutation takes over if the solution of the replicator equation tends to zero (${\mathcal G}(\xt;b_r,b_m)$ is negative), we define 
 \begin{eqnarray*} 
 	D(b_r) &:=&-\,\frac{\partial}{\partial b_m}{\mathcal G}(\xt;b_r,b_m)\bigg|_{b_m=b_r} 
 	= (b_r-C(b_r)\,\mu_s'(b_r)\,  
 	- C'(b_r) \mu_s(b_r)\nonumber
 \end{eqnarray*}
as the selection coefficient. A rare mutant with a trait arbitrary close, but larger than the resident's trait, invades if $D(b_r)>0$ (and for $D(b_r)<0$ we have the parallel conclusion).

\begin{theorem} Complete defection ($b=0$) is always an ESS. If $C'(0)>0$, complete defection ($b=0$) is a convergent stable ESS. If $C'(0)=0$ and 
$$\mu_s'(0)<-\mu_s(0)\,C''(0),$$ 
then the trait $b=0$ is an unstable ESS.
\end{theorem}
{\bf Proof: } We already know that $b=0$ is an ESS from Theorem~\ref{seedCoop}: The strategy ``always defect'' ($b=0$) can never be invaded by a rare mutant with $b>0$.\\
If $C'(0)>0$, then the selection coefficient at $b_r=0$ is negative, $D(0)=-C'(0)\mu_s(0)$, such that $b_r=0$ is a convergent stable ESS. If $C'(0)=0$ (an assumption that does not contradict that $C(b)$ is strictly increasing) then $D(0)=0$. The derivative of $D(b_r)$ at $b_r=0$ reads in this case
$$ D'(0) =-\mu_s'(0)-C''(0)\mu_s(0)$$
which is positive under the assumption $\mu_s(0)<-C''(0)\mu_s(0)$. Thus, for $b_r$ positive but small $D(b_r)$ is positive, and the ESS is unstable.
\par\qed\par\medskip 

We therefore find that cooperation can spread if $C'(0)=0$ and $\mu_s'(0)\ll 0$. If there is a slight tendency for cooperation to be present (for example introduced by stochastic fluctuations in a finite population, an aspect not considered here), there is a tendency for cooperation to invade the population or to increase in strength. The size of the seed bank produced by one plant is proportional to $\beta(b)/\mu_s(b)=\theta-\mu/\mu_s(b)$, and hence is decreasing if $\mu_s(.)$ decreases. In the present seed bank model, we conclude that cooperation correlates with less dormancy.

\section{Discussion}

The explicit formulation of life-history traits related to age structuring increases the state space of a model, leading to replicator equations that are more difficult to handle than the standard ones. In the first part of the paper, we show that the combination of two assumptions allow to return to a low dimensional replicator equation, that only is modified by the quiescent   and dormancy life-history traits. First, we assume that the life history traits do not come with a selective advantage or disadvantage. This assumption might always (by default) be satisfied as in the quiescence models, or need to be generated by specific parameter conditions as in the seed bank model. Second, we also assume that frequency-dependent selection is weak. This assumption is of course a restriction, but except for the cases of major genes of interaction underlying coevolution between symbionts, many frequency-dependent traits are likely polygenic and multi-locus so that single changes affect the trait and selection coefficient only weakly. Therefore, the second assumption covers a range of biologically interesting situations. Ultimately, the two assumptions have different effects. The first assumption leads to a transversally stable line of stationary points in a model with only two types and without frequency dependent selection (the procedure can be, of course, extended to more than two types). In the long run, the state of the system tends to this one-dimensional substructure. The second assumption, that is weak frequency-dependent selection, does not allow the system to move (far) away from that one-dimensional structure, but induces a slow flow on this one-dimensional structure, which yields the generalized replicator equation. Formally, singular perturbation theory is the method to derive our new replicator equation. The disadvantage of the present method compared with methods that are developed recently~\cite{Argasinski2017} is the loss of generality, but the advantage is the simple result that resembles the usual replicator equation.\par\medskip 

A main result is the difference between the resulting replicator equation for seed banks and quiescence. Specifically in the seed bank case, a certain condition on the parameters describing the seed germination is needed to ensure neutrality (equal fitness between types), while the quiescence models are generically neutral. At the root of this difference is the difference in the reproductive mechanism: Quiescent individuals are not involved in the reproduction, while in plants, reproduction happens via seeds. Indeed, in stationary populations, quiescence has no impact on the fitness of the types, we basically find the original replicator equation as a generalized replicator equation. We only need to take into account that the active individuals are under frequency-dependent selection and as such determine the replicator equation. Conversely, as seed banking and reproduction are strongly intertwined, in the seed bank model the seed germination time scales modify the replicator equation. \par\medskip 

In section 3, we use our theory to address an important problem in evolutionary theory: the stability of cooperation. We identify the frequency-dependent selection  term with the outcome of the prisoner's dilemma, and thus can study competition between cooperators and defectors with different life-history traits (quiescence or dormancy rates). For the quiescent models, and some conditions of the seed bank model, the life-history-traits of cooperators and defectors coincide, and we recover the well known tragedy of the commons (defector is an ESS). If, however, reproduction is strongly linked with the life-history trait as for seed banks, then it is possible that a population of only cooperators cannot be invaded by defectors (cooperation is an ESS). This result is in line with our previous study~\cite{Sellinger2019}, in which quiescence and seed banks were considered in exponentially growing populations. In that former case, quiescence influences the replicator equation, allowing for the stabilization of cooperation. We understand the difference between our two studies as follows. In exponentially growing populations, reproduction is hindered by quiescence, so that as a result, also  quiescence is strongly intertwined with reproduction. This is not the case in the current model with constant population size. Indeed, for the exponentially growing population, a condition regarding the cooperator and defector time scales (quiescence/dormancy rates) was required for cooperation to be maintained. This condition is similar to that of the current study in the seed bank model to establish neutrality. In other words, once population size is constant, quiescence does not promote cooperation, while seed banking can.\\ 
By means of adaptive dynamics, we show that defection always is an ESS, which can be for appropriate parameter functions become an unstable ESS. Namely, once a population moves away from the ``defector only'' strategy (e.g., by mutations and stochastic effects due to a finite population), then the level (strength) of  cooperation can increase and maintenance of cooperation occurs.\par\medskip 

From a technical point of view, we note that our approach does not only inherit two time scales, as it is usually the case for such models: the ecological time scale for the competition between a mutant and a resident, and the evolutionary time scale for the long term change of the trait. We require three time scales. On the fast time scale, the quiescent/active part of the population comes into an equilibrium. On a slower time scale, the principle of competitive exclusion yields the extinction of one of the two types. That time scale characterizes the ecological time scale. On a very slow time scale, rare mutants come in, and drive the resident's trait towards a convergent stable ESS. In general, it is thus an interesting question to understand how this evolutionary dynamics is affected by the life-history traits of the individuals (here quiescence and seed banking). \par\medskip 

One of our assumption requires some further discussion. We require a link/co-evolution between life history traits and cooperation. As suggested this link can be genetic pleiotropy with several genes (it has to be a multi-locus system underlying both quantitative traits) affecting dormancy (and seed germination) as well as cooperation. To our knowledge, there is to date no pleiotropic set of genes empirically studied. A second possibility, would be the existence of strong trade-offs between seed banking and cooperation, via possibly different sets of genes interacting at the phenotypic level. We are also not aware of such data, but we offer the hypothesis that such measures of degree of cooperation and germination rates can be assessed in natural populations. One can imagine to measure cooperation between plants in a field or between plants and their symbionts (\textit{e.g.} mycorrhizal fungi, nitrogen fixing bacteria), and assess their germination rates in different controlled conditions (of plants or symbionts). We are also not aware of studies assessing/linking empirically bacteria cooperation and rates of quiescence. From a modelling point of view, we observe that seed bank and quiescence models yield different results regarding the need for such pleiotropy / trade-offs is interesting but not fully understood by now. In conclusion, these findings show the potential of the general method to investigate the effect of life-history traits generating age-structuring on the replicator equation. As quiescence and dormancy are wide-spread and common features of plants \cite{tellier2019} and micro-organisms such as bacteria and fungi \cite{Lennon2021}, it is of interest to assess the co-evolution between these life-history traits and optimal evolutionary response to abiotic and biotic factors as well as with the stability and maintenance of cooperation or species trophic networks. We speculate that the increasing amount of information available for plant and animal microbiomes open the way to test our hypotheses and assess the role of quiescence/dormancy to maintain/promote cooperation.

\section*{Acknowledgements}
{\it {\bf Acknowledgements} This research is supported by a grant from the Deutsche Forschungsgemeinschaft (DFG) through TUM International Graduate School of Science and Engineering (IGSSE), GSC 81, within the project GENOMIE QADOP (JM+AT). AT areceives funding from the Deutsche Forschungsgemeinschaft (DFG) grant TE809/1-4, project 254587930.}

\bibliographystyle{alpha}
\bibliography{popGen,sleepyLit}

\newcommand{\etalchar}[1]{$^{#1}$}
\begin{thebibliography}{LdHWBB21}

\bibitem[AB17]{Argasinski2017}
Krzysztof Argasinski and Mark Broom.
\newblock Interaction rates, vital rates, background fitness and replicator
  dynamics: how to embed evolutionary game structure into realistic population
  dynamics.
\newblock {\em Theory in Biosciences}, 137(1):33--50, nov 2017.

\bibitem[AB21]{Argasinski2021}
Krzysztof Argasinski and Mark Broom.
\newblock Towards a replicator dynamics model of age structured populations.
\newblock {\em Journal of Mathematical Biology}, 82(5), apr 2021.

\bibitem[AN14]{AllenNowak2014}
Benjamin Allen and Martin Nowak.
\newblock Games on graphs.
\newblock {\em EMS Surv.\ Math.\ Sci.}, 1:113--151, 2014.

\bibitem[AS16]{AsfahlSchuster2016}
Kyle~L. Asfahl and Martin Schuster.
\newblock Social interactions in bacterial cell{\textendash}cell signaling.
\newblock {\em {FEMS} Microbiology Reviews}, 41(1):92--107, sep 2016.

\bibitem[BCKWB20]{Blath2020}
Jochen Blath, Adri{\'{a}}n~Gonz{\'{a}}lez Casanova, Noemi Kurt, and Maite
  Wilke-Berenguer.
\newblock The seed bank coalescent with simultaneous switching.
\newblock {\em Electronic Journal of Probability}, 25(none), jan 2020.

\bibitem[BGCKS13]{Blath2013}
Jochen Blath, Adri\'an Gonz\'alez~Casanova, Noemi Kurt, and Dario Span\`o.
\newblock The ancestral process of long-range seed bank models.
\newblock {\em J. Appl. Probab.}, 50:741--759, 2013.

\bibitem[BHS21]{Blath2021}
Jochen Blath, Felix Hermann, and Martin Slowik.
\newblock A branching process model for dormancy and seed banks in randomly
  fluctuating environments.
\newblock {\em Journal of Mathematical Biology}, 83(2), jul 2021.

\bibitem[BMC{\etalchar{+}}04]{Balaban2004}
Nathalie~Q Balaban, Jack Merrin, Remy Chait, Lukasz Kowalik, and Stanislas
  Leibler.
\newblock Bacterial persistence as a phenotypic switch.
\newblock {\em Science}, 305(5690):1622--1625, 2004.

\bibitem[{Die}04]{DiekmannIntro}
Odo {Diekmann}.
\newblock {A beginner's guide to adaptive dynamics.}
\newblock In {\em {Mathematical modelling of population dynamics. Collection of
  papers from the conference, B\c edlewo, Poland, June 24--28, 2002.}}, pages
  47--86. Warsaw: Polish Academy of Sciences, Institute of Mathematics, 2004.

\bibitem[DT96]{Day1996}
Troy Day and Peter~D. Taylor.
\newblock Evolutionarily stable versus fitness maximizing life histories under
  frequency-dependent selection.
\newblock {\em Proc. R. Soc. B}, 263:333--338, 1996.

\bibitem[GDdR05]{Getto2005}
Philipp Getto, Odo Diekmann, and Andre~M. de~Roos.
\newblock On the (dis) advantages of cannibalism.
\newblock {\em Journal of Mathematical Biology}, 51(6):695--712, nov 2005.

\bibitem[GVGC16]{Garay2016}
J{\'{o}}zsef Garay, Zolt{\'{a}}n Varga, Manuel G{\'{a}}mez, and Tomas Cabello.
\newblock Sib cannibalism can be adaptive for kin.
\newblock {\em Ecological Modelling}, 334:51--59, aug 2016.

\bibitem[HKM19]{Hoesel2019}
Volker H\"osel, Christina Kuttler, and Johannes M\"uller.
\newblock {\em Population Genetics and Bacterial Cooperation}.
\newblock {WORLD} {SCIENTIFIC}, 2019.

\bibitem[HMT{\v{Z}}18]{Heinrich2017}
Lukas Heinrich, Johannes M{\"u}ller, Aur{\'{e}}lien Tellier, and Daniel
  {\v{Z}}ivkovi{\'{c}}.
\newblock Effects of population- and seed bank size fluctuations on neutral
  evolution and efficacy of natural selection.
\newblock {\em Theoretical Population Biology}, 123:45--69, 2018.

\bibitem[HS98]{Hofbauer1998}
Josef Hofbauer and Karl Sigmund.
\newblock {\em Evolutionary games and population dynamics}.
\newblock Cambridge university press, 1998.

\bibitem[KKL01]{kaj2001}
Ingemar Kaj, Stephen~M. Krone, and Martin Lascoux.
\newblock Coalescent theory for seed bank models.
\newblock {\em J. Appl. Probab.}, 38:285--300, 2001.

\bibitem[KMT{\v{Z}}17]{Koopmann2017}
Bendix Koopmann, Johannes M{\"u}ller, Aur{\'e}lien Tellier, and Daniel
  {\v{Z}}ivkovi{\'c}.
\newblock Fisher--{W}right model with deterministic seed bank and selection.
\newblock {\em Theoretical population biology}, 114:29--39, 2017.

\bibitem[Kus05]{Kussell2005}
Edo Kussell.
\newblock Phenotypic diversity, population growth, and information in
  fluctuating environments.
\newblock {\em Science}, 309(5743):2075--2078, sep 2005.

\bibitem[LdHWBB21]{Lennon2021}
Jay~T. Lennon, Frank den Hollander, Maite Wilke-Berenguer, and Jochen Blath.
\newblock Principles of seed banks and the emergence of complexity from
  dormancy.
\newblock 12(1), aug 2021.

\bibitem[LGBT15]{Li2015}
Xiang-Yi Li, Stefano Giaimo, Annette Baudisch, and Arne Traulsen.
\newblock Modeling evolutionary games in populations with demographic
  structure.
\newblock {\em Journal of Theoretical Biology}, 380:506--515, sep 2015.

\bibitem[LS17]{Lessard2017}
Sabin Lessard and C{\'{\i}}ntia~Dalila Soares.
\newblock Frequency-dependent growth in class-structured populations:
  continuous dynamics in the limit of weak selection.
\newblock {\em Journal of Mathematical Biology}, 77(1):229--259, dec 2017.

\bibitem[Max14]{maxima}
Maxima.
\newblock Maxima, a computer algebra system. version 5.34.1, 2014.

\bibitem[McN13]{McNamara2013}
John~M. McNamara.
\newblock Towards a richer evolutionary game theory.
\newblock {\em Journal of The Royal Society Interface}, 10(88):20130544, nov
  2013.

\bibitem[MGM{\etalchar{+}}96]{Metz1995}
J.~A.~J. Metz, S.~A.~H. Geritz, G.~Meszéna, G.~Meszena, F.~J.~A. Jacobs, and
  J.S. van Heerwaarden.
\newblock Adaptive dynamics: A geometrical study of the consequences of nearly
  faithful reproduction.
\newblock In S.J. van Strien and S.M.~Verduyn Lunel, editors, {\em Stochastic
  and Spatial Structures of Dynamical Systems}, volume~45 of {\em KNAW
  Verhandelingen, Afd. Natuurkunde}, Amsterdam, 1996. North Holland.

\bibitem[MHF{\etalchar{+}}13]{Mueller2013}
Johannes M\"uller, Burkhard~A. Hense, Thilo~M. Fuchs, Margarete Utz, and
  Christian P\"otzsche.
\newblock Bet-hedging in stochastically switching environments.
\newblock {\em J. Theor. Biol.}, 336:144--157, nov 2013.

\bibitem[Oka09]{okasha2009}
Samir Okasha.
\newblock {\em Evolution and the Levels of Selection}.
\newblock Oxford University Press, 2009.

\bibitem[O'M12]{OMalley2012}
Robert O'Malley.
\newblock {\em Singular {P}erturbation {M}ethods for {O}rdinary {D}ifferential
  {E}quations}.
\newblock Springer New York, 2012.

\bibitem[Pri70]{price}
Georg~R. Price.
\newblock Selection and covariance.
\newblock {\em Nature}, 227:520--521, 1970.

\bibitem[SMHT19]{Sellinger2019}
Thibaut Sellinger, Johannes Müller, Volker Hösel, and Aur{\'{e}}lien Tellier.
\newblock Are the better cooperators dormant or quiescent?
\newblock {\em Mathematical Biosciences}, 318:108272, dec 2019.

\bibitem[TB09]{tellier2009}
Aurélien Tellier and James K.~M. Brown.
\newblock The influence of perenniality and seed banks on polymorphism in
  plant-parasite interactions.
\newblock {\em Am.\ Nat.}, 174:769--779, 2009.

\bibitem[Tel19]{tellier2019}
Aur{\'{e}}lien Tellier.
\newblock Persistent seed banking as eco-evolutionary determinant of plant
  nucleotide diversity: novel population genetics insights.
\newblock 221(2):725--730, sep 2019.

\bibitem[VT18]{verin2018host}
Melissa Verin and Aurelien Tellier.
\newblock Host-parasite coevolution can promote the evolution of seed banking
  as a bet-hedging strategy.
\newblock {\em Evolution}, 72(7):1362--1372, 2018.

\bibitem[WG10]{WestGardner2010}
Stuart~A West and Andy Gardner.
\newblock Altruism, spite, and greenbeards.
\newblock {\em Science}, 327(5971):1341--1344, mar 2010.

\bibitem[WGG07]{West2007a}
Stuart~A. West, Ashleigh~S. Griffin, and Andy Gardner.
\newblock Evolutionary explanations for cooperation.
\newblock {\em Current Biology}, 17(16):R661--R672, aug 2007.

\end{thebibliography}

\begin{appendix}
	
\newpage
\section{Transformations for models Quiescence II and Seedbank}

{\bf Proof} [of Proposition~\ref{quiesceceIItrafo}]\\
\begin{eqnarray*}
	\xt' &=& x' = 
	- x\,w\,\zeta_B(\mu\,+\eta_A)
	+(1-x)\,y\, \zeta_A( \mu+\eta_B) 
	+x\,(1-x)\,\beta (\eta_B-\eta_A)\\
	&&\qquad\qquad  + \eps\,\,\,x(1-x)\, ( g_A (\mu+\eta_B)- g_B(\mu+\eta_A))\\
	&=&		- x\,((1-x)\frac{\beta\,\eta_B}{\mu\,\zeta_B}+\eps \wt)\,\zeta_B(\mu\,+\eta_A)
	+(1-x)\,(x\,\frac{\beta\, \eta_A}{\mu\,\zeta_A}+\eps\yt)\, \zeta_A( \mu+\eta_B) \\
	&&+x\,(1-x)\,\beta (\eta_B-\eta_A)
	+ \eps\,\,\,x(1-x)\, ( g_A (\mu+\eta_B)- g_B(\mu+\eta_A))
	\\
	&=& \eps \bigg(
	- x\,\wt\,\zeta_B(\mu\,+\eta_A)
	+(1-x)\,\yt\, \zeta_A( \mu+\eta_B)
	+x(1-x)\, ( g_A (\mu+\eta_B)- g_B(\mu+\eta_A))\bigg).
\end{eqnarray*}

\begin{eqnarray*}
	\eps\yt' &=&  y'-x'\,\frac{\beta\, \eta_A}{\mu\,\zeta_A}\\
	&=&
	- (x\,\frac{\beta\, \eta_A}{\mu\,\zeta_A}+\eps \yt)\,\,\mu\zeta_A
	+x\,\,\beta \eta_A
	+x\, ((1-x)\,\frac{\beta\,\eta_B}{\mu\,\zeta_B}+\eps \wt)\,\, \eta_A  \zeta_B \\
	&&- (1-x) \,(x\,\frac{\beta\, \eta_A}{\mu\,\zeta_A}+\eps \yt) \,\,\eta_B \zeta_A +\eps \,\,\,\eta_A   \, ( g_A\, x+ g_B\,(1-x))\,x\\
	&&-\frac{\beta\, \eta_A}{\mu\,\zeta_A}\bigg(
	- x\,((1-x)\,\frac{\beta\,\eta_B}{\mu\,\zeta_B}+\eps \wt)\,\zeta_B(\mu\,+\eta_A)
	+(1-x)\,(x\,\frac{\beta\, \eta_A}{\mu\,\zeta_A}+\eps \yt)\, \zeta_A( \mu+\eta_B) \\
	&&+x\,(1-x)\,\beta (\eta_B-\eta_A)
	+ \eps\,\,\,x(1-x)\, ( g_A (\mu+\eta_B)- g_B(\mu+\eta_A))\bigg)\\
	&=&
	\eps\bigg(
	-\mu\zeta_A\, \yt 
	+\eta_A\zeta_B\,\xt\,\wt
	- \eta_B\zeta_A (1-\xt)\,\yt
	+ \eta_A   \, ( g_A\, \xt+ g_B\,(1-\xt))\,\xt\\
	&&\qquad 
	+\frac{\beta\eta_A\zeta_B(\mu+\eta_A)}{\mu\zeta_A}\, \xt\, \wt 
	-\frac{\beta\eta_A\zeta_A(\mu+\eta_B)}{\mu\zeta_A}\, (1-\xt)\, \yt \\
	&&\qquad 
	-\frac{\beta\, \eta_A}{\mu\,\zeta_A}
	x(1-x)\, ( g_A (\mu+\eta_B)- g_B(\mu+\eta_A))
	\bigg)\\
	&=&
	\eps\bigg(
	-\mu\zeta_A\, \yt 
	+\eta_A\zeta_B\,
	\frac{\beta(\mu+\eta_A)+\mu\zeta_A}{\mu\zeta_A}
	\xt\,\wt
	- \frac{\beta\eta_A(\mu+\eta_B)+\mu\eta_B\zeta_A }{\mu} (1-\xt)\,\yt\\
	&&\qquad +
	\eta_A   \, ( g_A\, \xt+ g_B\,(1-\xt))\,\xt
	-\frac{\beta\, \eta_A}{\mu\,\zeta_A}
	x(1-x)\, ( g_A (\mu+\eta_B)- g_B(\mu+\eta_A))
	\bigg)
\end{eqnarray*}

\begin{eqnarray*}
	\eps\wt' &=&  w'-(1-x)'\,\frac{\beta\, \eta_B}{\mu\,\zeta_B}\\
	&=&
	-((1-x)\,\frac{\beta\,\eta_B}{\mu\,\zeta_B}+\eps \wt)\,\,\mu \zeta_B
	+ (1-x) \,\,\beta  \eta_B
	+(1-x)\,(x\,\frac{\beta\, \eta_A}{\mu\,\zeta_A}+\eps \yt)\,\,\eta_B \zeta_A\\
	&&- x \, ((1-x)\,\frac{\beta\,\eta_B}{\mu\,\zeta_B}+\eps \wt) \,\,\eta_A \zeta_B 
	+ \eps\,\,\, 	\eta_B (g_A x+g_B (1-x))\, (1-x)\\
	&&+\frac{\beta\, \eta_B}{\mu\,\zeta_B}\bigg(
	- x\,((1-x)\,\frac{\beta\,\eta_B}{\mu\,\zeta_B}+\eps \wt)\,\zeta_B(\mu\,+\eta_A)
	+(1-x)\,(x\,\frac{\beta\, \eta_A}{\mu\,\zeta_A}+\eps \yt)\, \zeta_A( \mu+\eta_B) \\
	&&+x\,(1-x)\,\beta (\eta_B-\eta_A)
	+ \eps\,\,\,x(1-x)\, ( g_A (\mu+\eta_B)- g_B(\mu+\eta_A))\bigg)\\
	&=&\eps\bigg(
	-\mu\zeta_B\, \wt 
	-
	\frac{\beta\eta_B(\mu+\eta_A)+\mu\eta_A\zeta_B}
	{\mu}
	\xt\,\wt
	+\eta_B\zeta_A \frac{\beta(\mu+\eta_B)+\mu\zeta_B }{\mu\zeta_B} (1-\xt)\,\yt\\
	&&\qquad +
	\eta_B   \, ( g_A\, \xt+ g_B\,(1-\xt))\,(1-\xt)
	+\frac{\beta\, \eta_B}{\mu\,\zeta_B}
	x(1-x)\, ( g_A (\mu+\eta_B)- g_B(\mu+\eta_A))
	\bigg)\\
\end{eqnarray*}
\par\qed\par\medskip 

{\bf Proof:} [of Proposition~\ref{seedTrafo}]
\begin{eqnarray*}
	\frac d {dt} \tilde x 
	&=& 
	-\mu \tilde x 
	+ 
	\mu\,\frac{\theta \tilde x+\eps\,\tilde y}{\theta \tilde x+\eps\,\tilde y+z+\theta (1-x)+\eps\,\tilde z}	\\
	&=& 
	\mu\,\eps\,\frac{\tilde y\,(1-\tilde x)\,-\,\tilde z\,\tilde x}{\theta +\eps\,(\tilde y+\tilde z)}.	
\end{eqnarray*}

Recall that $\theta\,\mu_A = \beta_A-\mu$, s.t. 
\begin{eqnarray*}
	\eps\frac{d}{dt}\tilde y 
	&=& \frac d {dt} y- \theta\,\frac d {dt}\tilde x(t)\\
	&=& 
	-\, \mu\,\frac{y}{y+z}+ (\beta_A+ \eps\,g_A(x))x-\mu_Ay
	- \theta\,\frac d {dt}\tilde x(t)\\
	&=& 
	-\, \mu\,\frac{\theta \tilde x+\eps\,\tilde y}
	{\theta \tilde x+\eps\,\tilde y+\theta (1-\tilde x)+\eps\,\tilde z}
	+ (\beta_A+ \eps\,g_A(\tilde x))\tilde x-\mu_A(\theta \tilde x+\eps\,\tilde y)\\
	&& - \theta\,\mu\,\eps\,\frac{\tilde y\,(1-\tilde x)\,-\,\tilde z\,\tilde x}{\theta +\eps\,(\tilde y+\tilde z)}\\
	&=& 
	-\, \mu\,\frac{\theta \tilde x+\eps\,\tilde y}
	{\theta +\eps\,(\tilde y+\tilde z)}
	+\mu\tilde x+ \eps\,g_A(\tilde x)\tilde x-\eps \mu_A \,\tilde y
	- \theta\,\mu\,\eps\,\frac{\tilde y\,(1-\tilde x)\,-\,\tilde z\,\tilde x}{\theta +\eps\,(\tilde y+\tilde z)}
\end{eqnarray*}
\begin{eqnarray*}
	&=& 
	-\, \mu\,\frac{\theta \tilde x+\eps\,\tilde y-\tilde x(\theta +\eps\,(\tilde y+\tilde z))}
	{\theta +\eps\,(\tilde y+\tilde z)}
	+ \eps\,g_A(\tilde x)\tilde x-\eps \mu_A \,\tilde y
	- \theta\,\mu\,\eps\,\frac{\tilde y\,(1-\tilde x)\,+\,\tilde z\,\tilde x}{\theta +\eps\,(\tilde y+\tilde z)}\\
	&=& 
	-\, \eps\,\mu\,\frac{\tilde y(1-\tilde  x)- \tilde x\,\tilde z}
	{\theta +\eps\,(\tilde y+\tilde z)}
	+ \eps\,g_A(\tilde x)\tilde x-\eps \mu_A \,\tilde y
	+ \theta\,\mu\,\eps\,\frac{\tilde y\,(1-\tilde x)\,+\,\tilde z\,\tilde x}{\theta +\eps\,(\tilde y+\tilde z)}\\
	&=& 
	-\, \eps\,\mu\,(1+\theta)\,\frac{\tilde y(1-\tilde  x)- \tilde x\,\tilde z}
	{\theta +\eps\,(\tilde y+\tilde z)}
	+ \eps\,g_A(\tilde x)\tilde x-\eps \mu_A \,\tilde y.
\end{eqnarray*}
Dividing this equation by $\eps$ yields $\tilde y'(t)$. For $\tilde z(t)$, we obtain in a similar manner (using  $\theta\,\mu_B=\beta_B-\mu$)
\begin{eqnarray*}
	\eps\frac{d}{dt}\tilde z 
	&=& \frac d {dt} z+ \theta\,\frac d {dt}\tilde x\\
	&=&  -\, \mu\,\frac{z}{y+z}\,+ (\beta_B+\eps\,g_B(x))(1-x)-\mu_Bz 
	+ \theta\,\frac d {dt}\tilde x(t)\\
	&=&  -\, \mu\,\frac{\theta (1-\tilde x)+\eps\,\tilde z}{\theta \tilde x+\eps\,\tilde y+\theta (1-\tilde x)+\eps\,\tilde z}\,
	+ (\beta_B+\eps\,g_B(\tilde x))(1-\tilde x)-\mu_A(\theta (1-\tilde x)+\eps\,\tilde z) \\
	&&+ \theta\, \mu\,\eps\,\frac{\tilde y\,(1-\tilde x)\,-\,\tilde z\,\tilde x}{\theta +\eps\,(\tilde y+\tilde z)}\\
	&=&  -\, \mu\,\frac{\theta (1-\tilde x)+\eps\,\tilde z}{\theta +\eps\,(\tilde y+\tilde z)}\, 
	+\mu (1-\tilde x)
	+ \eps\,g_B(\tilde x)(1-\tilde x)-\eps\,\mu_B\,\tilde z 
	+ \theta\, \mu\,\eps\,\frac{\tilde y\,(1-\tilde x)\,-\,\tilde z\,\tilde x}{\theta +\eps\,(\tilde y+\tilde z)}\\
	&=&  -\, \mu\,\frac{\theta (1-\tilde x)+\eps\,\tilde z
		-  (1-\tilde x) (\theta +\eps\,(\tilde y+\tilde z))
	}{\theta +\eps\,(\tilde y+\tilde z)}\, \\
	&&+ \eps\,g_B(\tilde x)(1-\tilde x)-\eps\,\mu_B\,\tilde z 
	+ \theta\, \mu\,\eps\,\frac{\tilde y\,(1-\tilde x)\,-\,\tilde z\,\tilde x}{\theta +\eps\,(\tilde y+\tilde z)}\\
	&=&  \, \mu\,\eps\,\frac{
		(1-\tilde x)\,\tilde y-\tilde x \tilde z
	}{\theta +\eps\,(\tilde y+\tilde z)}\, 
	+ \eps\,g_B(x)(1-\tilde x)-\eps\,\mu_B\,\tilde z 
	+ \theta\, \mu\,\eps\,\frac{\tilde y\,(1-\tilde x)\,-\,\tilde z\,\tilde x}{\theta +\eps\,(\tilde y+\tilde z)}\\
	&=&  \, \mu\,\eps\,(1+\theta)\,\frac{\tilde y\,(1-\tilde x)\,-\,\tilde z\,\tilde x}{\theta +\eps\,(\tilde y+\tilde z)}\, 
	+ \eps\,g_B(x)(1-\tilde x)-\eps\,\mu_B\,\tilde z 
\end{eqnarray*}
\par\qed
\par\medskip

\end{appendix}

\end{document}